# Influence of vacancies on phonons
# and pressure phase transitions in TiO: a DFT study

S. Vahid Hosseini[1]; Mohaddeseh Abbasnejad[2];

Mohammad Reza Mohammadizadeh[1*],

[1] *Superconductivity Research Laboratory (SRL), Department of Physics, University of Tehran, North Kargar Ave., P. O. Box 14395-547, Tehran, Iran.*

[2] *Faculty of Physics, Shahid Bahonar University of Kerman, Kerman, Iran*

**\*Corresponding Author:**

**Mohammad Reza Mohammadizadeh,**

**Prof. of Physics,**

**Permanent address:**

**Director, Superconductivity and Supermaterials Research Laboratory (SRL),**

**Department of Physics, University of Tehran,**

**North Kargar Ave.,**

**P.O. Box 14395-547**

**Tehran,**

**IRAN**

**Tel: +98 21 61118611, 61118634, & 61118749**

**Fax: +98 21 88004781**

zadeh@ut.ac.ir

# Influence of vacancies on phonons
# and pressure phase transitions in TiO: a DFT study

S. Vahid Hosseini[1]; Mohaddeseh Abbasnejad[2];

Mohammad Reza Mohammadizadeh[1*],

[1] *Superconductivity Research Laboratory (SRL), Department of Physics, University of Tehran, North Kargar Ave., P. O. Box 14395-547, Tehran, Iran.*

[2] *Faculty of Physics, Shahid Bahonar University of Kerman, Kerman, Iran*

## Abstract

Since recently, some interests appeared in TiO as a thin film coating material due to its promising physical properties. Moreover, because of the existence of intrinsic vacancies, TiO demonstrates memristive properties. To use these compounds in active region for memristive switching, the investigation of the dynamical stability of these compounds is of importance. Therefore, the structural, electronic, and dynamical properties of TiO in two different phases, namely cubic and monoclinic structures along with vacancies, are debated from the framework of density functional theory. The structural calculations show that the vacancies have quite an impact on the bulk modulus of these compounds. The electronic band structure remarks that these structures are metallic. In addition, there is an energy gap between O-$2p$ and Ti-$3d$ states in these compounds below the Fermi level that can be altered via vacancy concentration. The phonon calculations reveal that monoclinic structure is dynamically stable whereas the cubic phase is unstable at room temperature. Considering ~12.5% of both Ti and O vacancy concentration, eliminate the imaginary modes for the cubic structure. Moreover, the vacancy-free cubic structure can be stable above ~ 10 GPa applied pressure, which is consistent with the experiment. Additionally, phonon dispersion curves of $TiO_x$ compounds suggest that these materials are not favorable for thermal conductivity. The pressure phase transition results indicate the transition from monoclinic to vacant cubic TiO at ~8 GPa, whereas the monoclinic to vacancy-free cubic TiO transition occurs at ~ 44 GPa, which are in a good agreement with the recent reports. Therefore, it can be concluded that engineering of the vacancy sites and their concentration in TiO phases can play an important role in applying them in technological applications.

# 1 Introduction

Nowadays, Ti-O based devices are widely used in industrial applications [1-4]. Accordingly, understanding the various properties of Ti-O systems is crucially important. Recently, Li *et al.* [5] have succeeded in predicting 27 compounds of Ti-O system ($Ti_yO_x$ compounds) using a combination of variable-composition evolutionary crystal structure prediction and data mining in the pressure range 0−200 GPa. The other widely known Ti-O systems are titanium suboxide with $Ti_nO_{2n-1}$ (n≥1) chemical formula. Titanium monoxide (TiO) is an inorganic chemical material in this family, comprising two different phases: (i) The cubic phase (similar to that of NaCl) having symmetry of $Fm\bar{3}m$, which is synthesized at a high-temperature (1400 ℃). In this phase, there are ~14.6 % randomly arranged vacancies on both titanium and oxygen sublattices [6]. (ii) The monoclinic phase with *A2/m* symmetry that is more stable than the cubic one and contains 5 TiO units in its unit cell. In fact, the conventional unit cell of the monoclinic TiO contains twelve Ti and twelve O sites in which one sixth of both sites are assumed to be vacant, i.e. two titanium and two oxygen atoms are missing from the unit cell [6]. This phase is structured at a lower-temperature via the rapid cooling from 1400 to 990 ℃ such that the vacancies are ordered [7]. It should be added that the samples of the vacancy-free rock-salt structure can be obtained by annealing monoclinic phase at high temperature and pressure [8]. The stoichiometric TiO has a metallic conductivity and shows superconductivity properties. Moreover, this structure has different applications as a thin film coating material or diffusion barrier against the interdiffusion of Al and Si [9,10]. However, its physical properties may alter due to its non-stoichiometric compounds with $TiO_x$ formula. In fact, $TiO_x$ can appear in diverse compositions where x ranges from 0.7 to 1.25 [7]. The vacancy concentration can have an impact on the conductivity and superconductivity properties of TiO [11]. For instance, Hulm *et al*. [12] reported $T_C$ ~ 1 K for the cubic $TiO_{1.07}$. Later, Wang *et al*. [13] could reach $T_C$ ~ 5.5 K by setting the exact Ti/O molar ratio of 1:1. Additionally, the available vacancies introduce $TiO_x$ as a new and promising candidate in memristors for synaptic device applications [14-16]. To explain this, when a voltage is applied to the vacant TiO structure, the ordered and random vacancies start moving through the material.  As the voltage is switched, the vacancies are not able to have an initial re-arrangement in the structure. Thereafter, the material shows different resistance in a voltage switching. This results in appearance of resistance hysteresis in I-V diagram. Thus, vacant TiO can be fascinating for the active medium technology [17,18]. Regarding that, it is important to find out the role of intrinsic defects in mechanical and dynamical stability of this structure. Although, there are a few reports focusing on the electrostatic origin of stability by vacancies in TiO, there is no report on the dynamical stability of TiO$_x$, up to now. Furthermore, the monoclinic TiO can experience structural phase transition into phase by applying the temperature or pressure. It has been reported the transition pressure accrues at ~ 7.7 GPa while cubic phase contains 14.4% random vacancies, experimentally [19]. However, Leung *et al.* [7] predicted this transition pressure ~25 GPa. On the other hand, Ding *et al.* [20] remarked that the monoclinic phase goes through the cubic one at 37.8 GPa, when the cubic phase is defect-free. Here, we confirm our findings to reproduce these experimental phase transitions.

Therefore, from what has been discussed, this paper aims to address the influence of available vacancies in the structural and electronic properties of TiO phases. In addition, it is attempted to investigate the effects of vacancies on phonons and dynamical stability of these compounds. Finally, the present work tries to acknowledge the reports on the pressure phase transition of these materials. The present work is organized as follows. Section 2 contains computational details. In section 3, the results and discussion over the topics as structural and electronic properties, the lattice dynamics and phonon properties, and prediction of pressure phase transition of TiO phases are debated. The conclusions are summarized in section 4.

## 2 Computational details

The calculations were performed in the framework of density functional theory (DFT) using the Quantum ESPRESSO package with plane-wave basis set and pseudopotentials method[21]. In the present calculations, the ultrasoft pseudopotentials were applied to describe the interaction between electrons and ions [22]. The 3s, 3$p$, 3$d$, and 4$s$ electrons (2$s$ and 2$p$ electrons) for Ti (O) were treated as valence, and the remaining electrons were fixed frozen in the calculations. The generalized gradient approximation (GGA) with the flavor of Perdew-Burk-Ernzerhoff (PBE) functional was used to approximate the exchange-correlation interaction [23]. The kinetic cutoff energy for the plane wave and charge density expansion of wave functions for the cubic (monoclinic) phases were opted as 45 (60) and 450 (720) Ry, respectively. The energy convergence was set to $10^{-6}$ Ry. The Brillouin zone for both the cubic and monoclinic phases was sampled applying the Monkhorst-Pack scheme [24] with k-point grids of 20×20×20 (equivalent to 256 k-points in the irreducible first Brillouin zone (1$^{st}$ BZ)) and 8×8×8 (equivalent to 260 k-points in the irreducible $1^{st}$ BZ), respectively. The Broyden–Fletcher–Goldfarb–Shanno (BFGS) algorithm was applied to gain a full geometry relaxation [25]. In the geometry optimization, all the coordinates of the atoms were relaxed until the Hellman–Feynman forces were less than $10^{-5}$ Ry/a.u. For the lattice dynamics investigations, the Phonopy package [26] based on Parlinski-Li-Kawazoe method [27] with a finite displacement of 0.01 Å was used to calculate the phonon dispersion curves and thermodynamic properties of the phases in question. It was found that $4\times4\times4$ and $2\times2\times2$ supercells are enough to fulfill the desired precision (the energy difference between the selected supercells is in the order of meV/unit cell) for the cubic and monoclinic phases, respectively.

## 3 Results and discussion

### 3.1 Structural and electronic properties

In this paper we have studied the nominal TiO compositions. The defect-free cubic and monoclinic TiO are denoted as c-TiO and m-TiO, respectively. The defected cubic $Ti_{7/8}O_{7/8}$, equivalently c-$Ti_{0.875}O_{0.875}$, includes 12.5 % of both Ti and O vacancies, whereas the defected monoclinic $Ti_{6/6}O_{5/6}$, equivalently m-$TiO_{0.833}$, has 16.6 % O vacancies, and the defected monoclinic $Ti_{5/6}O_{6/6}$, equivalently m-$TiO_{1.2}$, contains 16.6 % Ti defects. A schematic view of these nominal TiO compositions was represented in Fig. 1. It should be mentioned that there are several configurations for considering vacancy sites to construct the proposed vacant cubic TiO, i.e. c-$Ti_{0.875}O_{0.875}$. Out of 49 different configurations ($2\times2\times2$ supercell of c-TiO in which one Ti and O are removed), there were only 8 ones maintaining the cubic symmetry. For instance, considering the first neighbor Ti-O vacancy changed the symmetry to tetragonal (*I4mm* (107)). Therefore, we chose the structure with Ti and O removed at (0, 0, 0) and (0.5, 0.5, 0.5), respectively; which is energetically lower by the order of a few meV compared to the others. The same procedure was also repeated for the other structures containing vacancy sites. Furthermore, $Ti_{6/6}O_{5/6}$ constructed from the primitive cubic TiO which was transformed into the monoclinic space group after geometry relaxation.

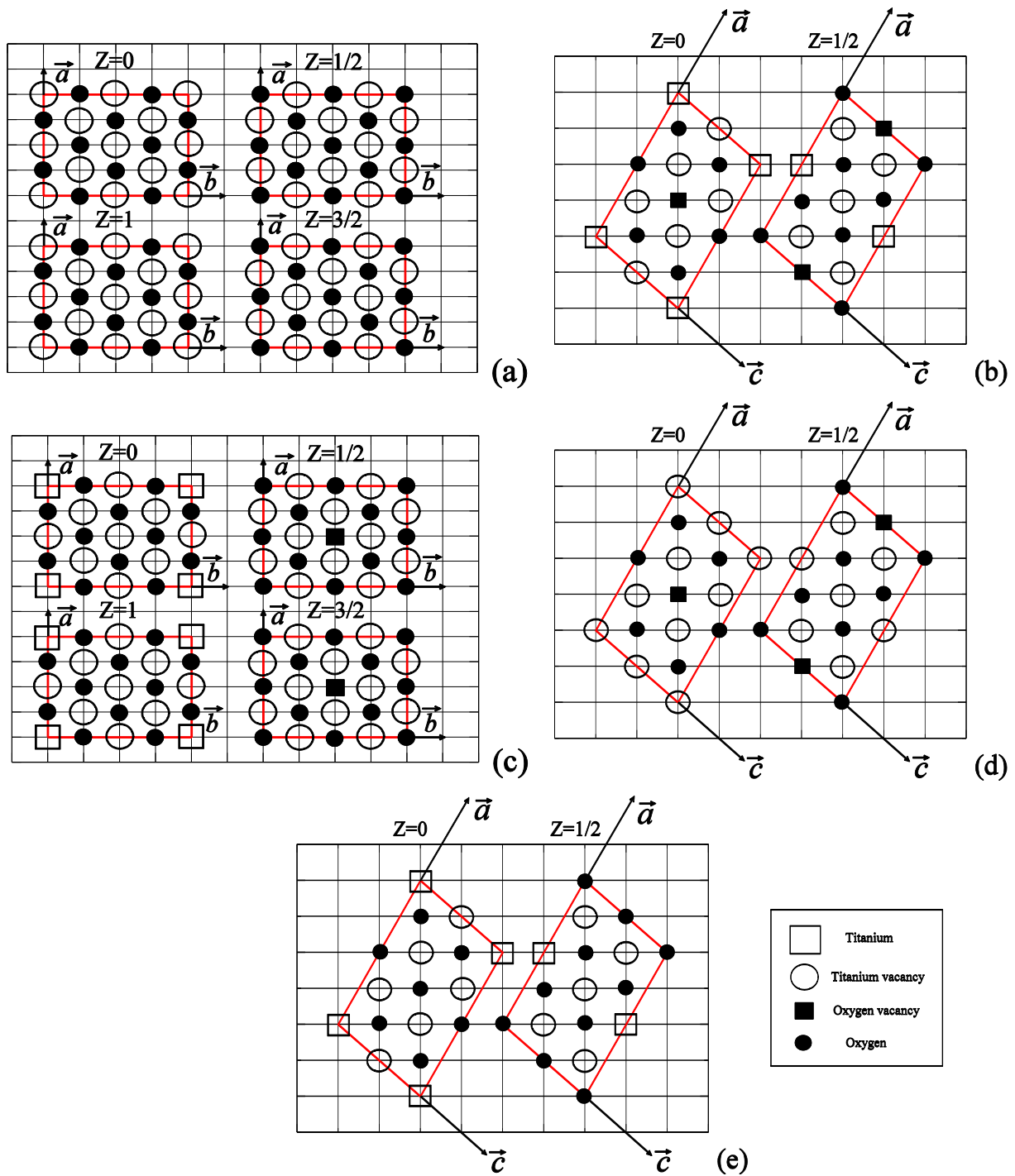

**Fig. 1.** Overall view of the TiO crystal unit cells (a) c-TiO, (b) m-TiO, (c) c-Ti$_{0.875}$O$_{0.875}$, (d) m-TiO$_{0.833}$, and (e) m-TiO$_{1.2}$. The large and small circles are representative of titanium and oxygen atoms, respectively. The dash balls are corresponding to Ti or O vacant atomic sites introduced into the system.

The crystal structure of c-TiO and m-TiO phases including lattice parameters, bulk modulus, the pressure derivative of bulk modulus, and the relaxed volume can be found in Table 1.

**Table 1.** The optimized structural parameters including lattice parameters ($a,b,c$), lattice angle ($\gamma$), equilibrium volume ($V_0$), bulk modulus ($B_0$), and pressure derivative of bulk modulus ($B_0{}'$) for c-TiO and m-TiO phases. The calculations were performed at zero pressure and temperature.

| Structure | This Work | Others | Exp. | Atoms | Wyckoff positions | Atomic positions (This work) | (Exp.)[6] |
|---|---|---|---|---|---|---|---|
| **Cubic** | | | | | | | |
| $a$/(Å) | 4.27 | 4.15 [28] | 4.18 [29] | | | | |
| $V$/(Å³) | 77.85 | 76.76 [28] | 73.03 [29] | | | | |
| $B_0$/(GPa) | 225.5 | 260 [30] | 200 [30] | | | | |
| $B_0{}'$ | 3.91 | 3.76 [30] | | | | | |
| **Monoclinic** | | | | | | | |
| $a$/(Å) | 5.84 | 5.73 [7] | 5.85 [6] | 2Ti | 2c | (0.50, 0.00, 0.00) | (0.50, 0.00, 0.00) |
| $b$/(Å) | 9.34 | 9.27 [7] | 9.34 [6] | | | (0.50, 0.50, 0.50) | (0.50, 0.50, 0.50) |
| $c$/(Å) | 4.17 | 4.13 [7] | 4.14 [6] | 2O | 2b | (0.00, 0.50, 0.00) | (0.00, 0.50, 0.00) |
| $\gamma$/(°) | 107°31´ | 107°00´[7] | 107°32´[6] | | | (0.00, 0.00, 0.50) | (0.00, 0.00, 0.50) |
| $V$/(Å³) | 217.59 | 209.78[7] | 215.94[6] | 4Ti | 4i | (0.16, 0.33, 0.00) | (0.15, 0.32, 0.00) |
| $B_0$/(GPa) | 197.5 | 250 [30] | | 4Ti | 4i | (0.66, 0.33, 0.00) | (0.66, 0.33, 0.00) |
| $B_0{}'$ | 3.55 | 3.17 [30] | | 4O | 4i | (0.33, 0.16, 0.00) | (0.33, 0.16, 0.00) |
| | | | | 4O | 4i | (0.83, 0.16, 0.00) | (0.82, 0.16, 0.00) |

The estimated error in the volume for c-TiO and m-TiO is 6.60% and 0.7%, respectively. This error is rather high for the cubic phase. This is because of ignorance of intrinsic vacancies in this structure. Including 12.5% vacancies in both Ti, and O, i.e. c-Ti$_{0.875}$O$_{0.875}$ corrects the lattice parameters to 4.17 Å. It shows that including the vacancies reduces the lattice parameter, which is more consistent with the experiment [8,30]. To have a better analysis with the experiment, the XRD pattern of c-TiO, and c-Ti$_{0.875}$O$_{0.875}$ taken from VESTA package [31] and the experiment [20] (extracted from 2θ= 4° to 22°) are presented in electronic supplementary information (ESI). According to Fig. S1 in the ESI, one can see that the experimental peaks shown by solid red circles observed in the lower degrees (2θ=14°-18°) are not produced in c-TiO. These peaks stand out of intrinsic vacancies in the cubic phase. The discrepancy in peaks' intensity between experiment and c-$Ti_{0.875}O_{0.875}$ originates from intrinsic disordering vacancies, which are not considered in our simulation. The detailed information about the atomic positions of c-$Ti_{0.875}O_{0.875}$ is given in Table S1 in the ESI. For the case of m-TiO, it turns out that the obtained structural results in GGA calculations give a better description compared to those reported in Ref. [7], which were performed in local density approximation (LDA). The calculated bulk modulus for m-TiO is less than that of c-TiO phase. It indicates that c-TiO phase is more resistant to compression relative to m-TiO one. The calculated bulk moduli in c-TiO and c-$Ti_{0.875}O_{0.875}$ phases are 225.5 and 188.7 GPa, respectively. For the case of monoclinic phases as m-TiO, m-$TiO_{0.833}$, and m-$TiO_{1.2}$, the bulk modulus was calculated 197.5, 173.4, and 130.3 GPa, respectively. It is clear that the bulk modulus is decreased by considering the vacancies for c-TiO and m-TiO phases. It implies that the vacancies in the cubic and monoclinic phases can make quite an impact on their bulk modulus. In this regard, there is no experimental data to compare the results.

In terms of the electronic properties, both materials are experimentally metallic. However, m-TiO conductance is higher than that of c-TiO phase and it increases more rapidly from the room temperature [29]. Figure 2 shows the band structure, total and partial electronic density of states (DOS) for both c-TiO and m-TiO phases. The bands cross the Fermi level indicating the metallic behavior of these phases. The partial DOS is manly made of O-2$p$ states bellow the Fermi level, and Ti-3$d$ at the Fermi level, whereas Ti-3$s$, 3$p$, and 4$s$ bands have almost zero contribution around the Fermi level. There is a $p$-$d$ gap region observed in both phases that can be altering by the vacancy concentration. The results of electronic calculations for c-TiO and m-TiO phases are in well agreement with Ref. [32] and Ref. [28], respectively. To gain more detail discussions in the electronic properties of TiO, there are already several reports that a reader can refer to Refs. [33-37].

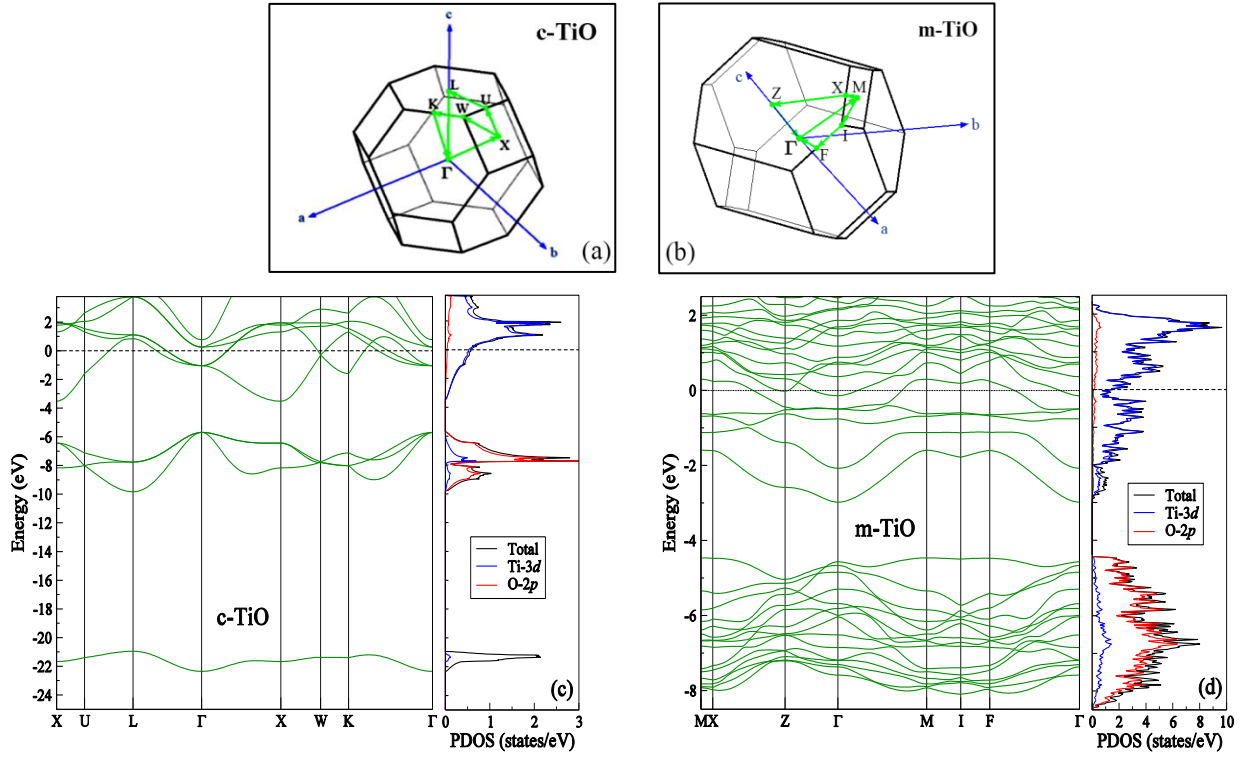

**Fig. 2.** The primitive Brillouin zone of (a) c-TiO, (b) m-TiO, (c) and (d) represents their corresponding band structure, total and partial electronic density of states, respectively. The Fermi level is set to zero.

The electronic DOS for the other compositions as c-Ti$_{0.875}$O$_{0.875}$, m-TiO$_{0.833}$, and m-TiO$_{1.2}$ are provided in ESI (Fig. 2S). The general shape of DOS follows the trends of c-TiO and m-TiO. The difference in the electronic DOS for these structures comes from the number of sharp peaks below the Fermi level originating from Ti-3$d$ states, the width of $d$ bands in accordance with vacancy concentration, and $p$-$d$ gap. Furthermore, the Fermi level is located on the peak for c-Ti$_{0.875}$O$_{0.875}$, while it was in the pseudogap region for m-TiO$_{0.833}$, and m-TiO$_{1.2}$.

### 3.2 Lattice dynamics and phonon properties

The phonon dispersion curves for c-TiO and m-TiO are depicted in Fig. 3 (a) and (b), respectively.

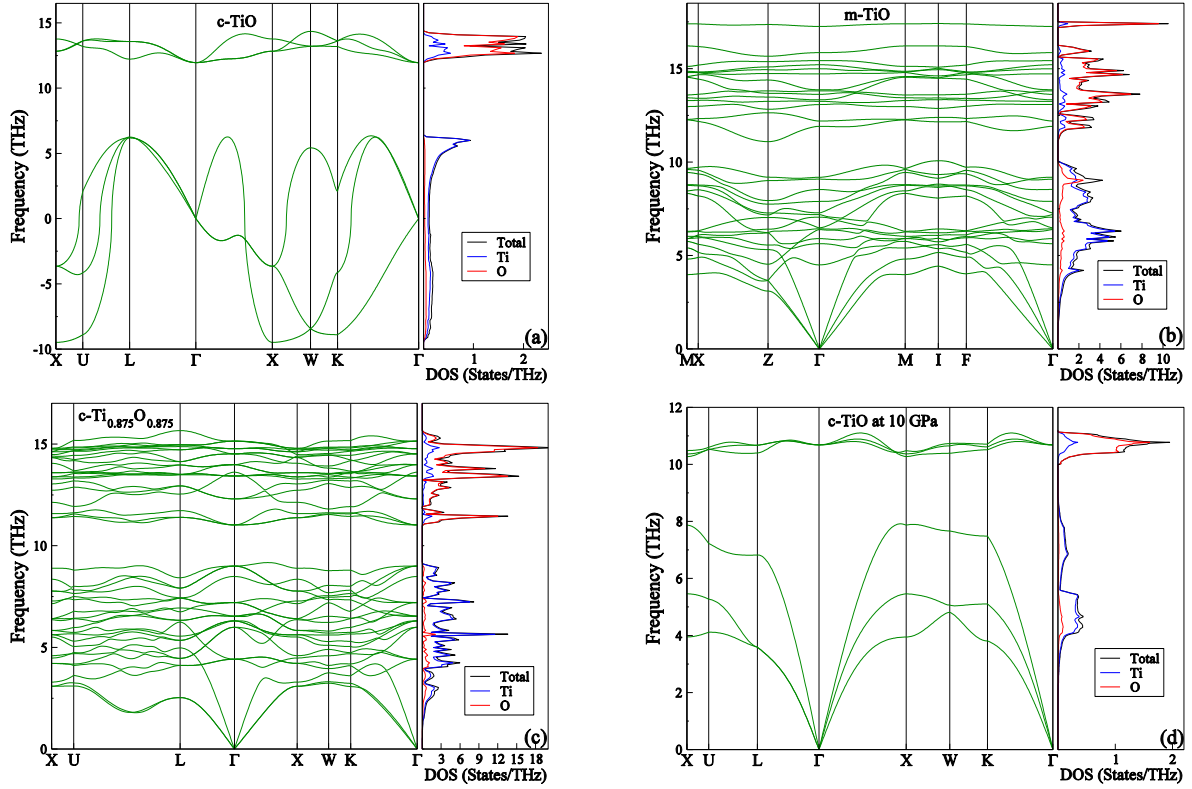

**Fig. 3.** The phonon dispersion curve and phonon density of states for TiO crystal phases (a) c-TiO, (b) , m-TiO (c) c-Ti$_{0.875}$O$_{0.875}$, and (d) c-TiO at 10 GPa.

The phonon dispersion curves display that c-TiO has three acoustic imaginary modes, whereas the m-TiO is dynamically stable. To investigate the unstable modes in c-TiO, the energy and force thresholds were checked carefully. In addition, different kinetic cut-off energy and pseudopotentials, smearing parameter ($\sigma$), some dense k-point samplings and larger supercells were tested. Additionally, this phase was examined by density functional perturbation theory approach [38] and a full-potential augmented plane wave method as implemented in WIEN2k [39]. However, the unstable modes were not eliminated. On the other side, the question that may arise is whether these unstable modes can be responsible for the structural phase transition into the monoclinic phase? To investigate this, the modulation is used to create a crystal structure with displacements along normal modes at q-point in the specified supercell dimension. Atomic displacement of the $j^{\text{th}}$ atom is created from the real part of the eigenvectors with amplitudes and phase factors as:

$$\frac{A}{\sqrt{N_a M_k}} Re[\exp(i\varphi)\, \boldsymbol{e}_k \exp(\boldsymbol{q} \cdot \boldsymbol{r}_{km})], \tag{1}$$

where $A$ is the amplitude, $\varphi$ is the phase factor, $N_a$ is the number of atoms in the supercell, $M_k$ is the mass of the $k^{\text{th}}$ atom, $\boldsymbol{q}$ is the q-point specified, $\boldsymbol{r}_{km}$ is the position of the $k^{\text{th}}$ atom in the $m^{\text{th}}$ unit cell, and $\boldsymbol{e}_k$ is the $k^{\text{th}}$ atom part of eigenvector. Thereafter, according to Fig. 3(a), X, W, and K points represent imaginary frequencies. X, W, and K are q-points corresponding to (0, 0.5, 0.5), (0.25, 0.5, 0.75) and (0.375, 0.375, 0.75) in the reciprocal lattice, respectively. Hence, Eq. (1) creates a supercell with $1\times2\times2$, $12\times6\times4$, and $8\times8\times4$ cell dimensions, respectively. Since the required supercell for W and K points are rather large for testing, the X point was considered. Table 2 shows the modulation with the phonon wave vector (X) through the solid for the imaginary band index with amplitude of unit and different phase factors [27].

**Table 2.** The modulation of cubic phase using $1\times2\times2$ supercell at q-point (0, 0.5, 0.5).

| Band index | Phase factor (°) | Symmetry results after relaxation |
|:---:|:---:|:---:|
| 1 | 0 | Orthorhombic, *Amm2* (38) |
| 1 | 90 | Orthorhombic, *Cmmm* (65) |
| 1 | 180 | Orthorhombic, *Amm2* (38) |
| 1 | mix | Orthorhombic, *Cmmm* (65) |
| 2 | 0 | Monoclinic, *Cm* (8) |
| 2 | 90 | Orthorhombic, *Cmmm* (65) |
| 2 | 180 | Monoclinic, *Cm* (8) |
| 2 | mix | Orthorhombic, *Cmmm* (65) |
| 3 | 0 | Monoclinic, *Cm* (8) |
| 3 | 90 | Orthorhombic, *Cmmm* (65) |
| 3 | 180 | Orthorhombic, *Cmcm* (63) |
| 3 | mix | Orthorhombic, *Cmmm* (65) |

From Table 2, only *Cm* (8) is a monoclinic phase, which is not observed in experiment (*A2/m*). However, we tested the dynamical stability of *Cm* (8) space group. The result of phonon dispersion curve revealed that this phase is dynamically unstable at room temperature.

From above discussion, it can be concluded that the vacancy-free rock-salt structure is not stable at room temperature. As discussed earlier, this phase includes ~15 % disordered vacancies on both Ti and O atoms, inherently. It is straightforward that considering vacancies (c-Ti$_{0.875}$O$_{0.875}$ shown in Fig. 3(c)) corrects the imaginary modes. As it was mentioned in section 3.1, there are 8 distinct structures for c-Ti$_{0.875}$O$_{0.875}$ with the same cubic symmetry. It is worthy to say that the phonon dispersion curve of these structures was not vastly affected by the position of Ti and O vacancy in these configurations. The results can be found in Fig. 3S in ESI. On the other hand, the experimental report remarks that above 8 GPa pressure, the vacancies can be eliminated from the vacant TiO and the vacancy-free sample is obtained. However, our phonon calculation predicted that c-TiO can be dynamically stable above 10 GPa pressure as displayed in Fig. 3(d). As it is observed c-TiO shows a phonon band gap of 2.3 THz. Furthermore, the optical modes are degenerate in $\Gamma$ point that is indicative of metallic character. The pressure has lifted the degeneracy for transverse acoustic modes in all directions except $\Gamma$-L in c-TiO. The flat optical modes create sharp phonon density of states that can be potential for superconductivity of these structures.

Comparison between the c-Ti$_{0.875}$O$_{0.875}$ and m-TiO dispersion curves illustrated in Fig. 3 shows that the degenerate acoustic modes are broken in the monoclinic phase. It can be due to Ti and O missing atoms that distort squares and hexagons in the 1$^{st}$ BZ of monoclinic structure (see Fig. 2(b)). There is no phonon gap between acoustic and optical modes for both phases. This suggests that the acoustic modes, which are responsible for heat transfer, can easily exchange energy with optical modes, thus resulting in low thermal conductivity in these compounds. Moreover, a phonon band gap is also observed in the m-TiO and c-Ti$_{0.875}$O$_{0.875}$ phase around 10 THz. However, this band-gap is not seen in NaCl structure [40], which its space group is the same as c-TiO. The origin of phonon band-gap may be traced to the big atomic mass ratio of Ti and O, which is ~3, while the Na/Cl mass ratio is about 1.54. This relatively big ratio in TiO crystals can open up a band-gap. In addition, the difference magnitude of phonon band gap in c-Ti$_{0.875}$O$_{0.875}$ and m-TiO compositions is attributed to non-equivalent bonds in these structures. The flat bands in higher frequencies can be due to the rigid-unit modes (RUMs) [41,42] in TiO crystals in which TiO$_6$ octahedra move together as a rigid unit and consecutively do not distort the bands connecting the atoms within the unit cell. Due to the rigidity of the motion, rather undispersed bands are observed in both phases above 10 THz phonon frequency ranges.

Figure 4 (a) shows phonon dispersion curves and phonon DOS for the m-TiO$_{0.833}$ and m-TiO$_{1.2}$ phases. As observed, acoustic modes of m-TiO$_{0.833}$ shift to higher energies (acoustic hardening) compared to m-TiO. Conversely, the acoustic modes shift to lower energies (acoustic softening) in m-TiO$_{1.2}$ with respect to m-TiO as represented in Fig. 4(b). This fact implies that introducing oxygen vacancies stabilizes monoclinic TiO more than the titanium vacancies, since the atomic force constant is stronger. Therefore, it is expected that TiO$_x$ thin films can be more stable in Ti-rich conditions.

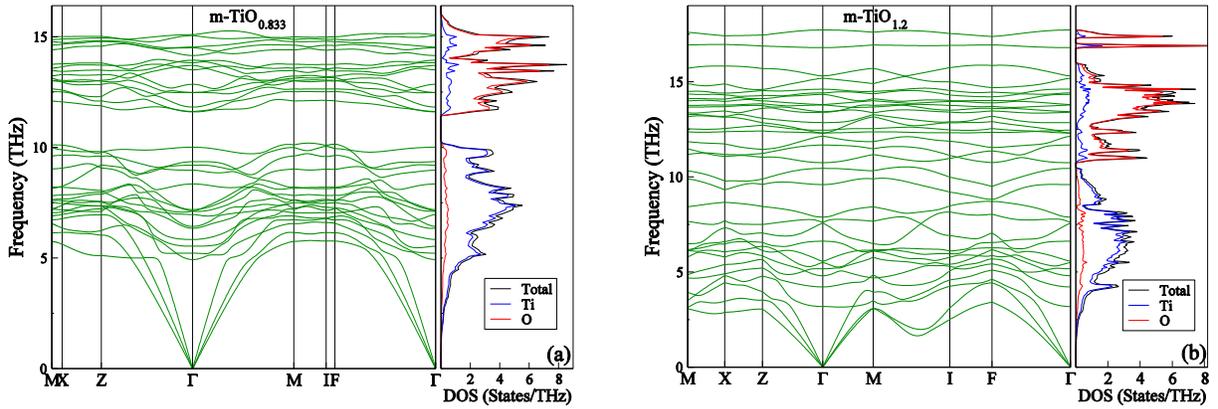

**Fig. 4.** The phonon dispersion curve and phonon density of states for TiO chemical compounds (a) m-TiO$_{0.833}$, and (b) m-TiO$_{1.2}$.

Generally, as it is viewed in phonon PDOS in all compositions, Ti atoms make a major contribution to lower frequencies, whereas O atoms play the role in the higher frequencies. This can be anticipated because Ti atomic mass is greater than that of O atom. On the other hand, one can conclude that the O atoms contribute more to the optical modes. For the m-TiO case, there is a sharp peak in the projected phonon DOS at ~17 THz (Fig. 3(b)), remarking localization of the states for the oxygen atoms, i.e., they treat like "atomic states" in accordance with a single flat band observed in its phonon dispersion curve. The single-band is disappeared by introducing oxygen vacancy in the monoclinic (i.e. m-TiO$_{0.833}$) and placed among other oxygen phonon states as shown in Fig. 4(a). Specifically speaking, there are two types of O atoms in the structure: One, which is bonded to 4 Ti atoms and the other, is 5-bonded. This single non-degenerate mode is related to an out-of-basal plane vibration of the 4-bonded oxygen atoms, while its in-plane vibrations ought to be 2-degenerated and placed in the frequency ranges between ~ 12-15 THz. In addition, in these frequency ranges, 5-bonded oxygen atoms form either in plane or out-of-plane vibrations at the same time that are appeared as the flat bands. To our knowledge, there are not adequate experimental reports corresponding to phonon dispersion curve of TiO phases. Table 3 shows the experimental and the calculated Raman frequencies (cm$^{-1}$) for TiO phases. The experimental data is not available for c-TiO phase. As it is seen, the optical modes of c-TiO are degenerate indicating the isotropic medium of this phase. The available experimental data and the calculated Raman modes (cm$^{-1}$) are rather in an agreement and the difference may come back to the shortcomings of GGA functional.

**Table 3.** The experimental and computed Raman frequencies (cm$^{-1}$) for TiO phases

| c-TiO | | m-TiO | | |
|---|---|---|---|---|
| Mode | This work (cm$^{-1}$) | Mode | This work (cm$^{-1}$) | Exp. [43] (cm$^{-1}$) |
| $T_{1u}(3)$ | 352.7 | $B_u$ | 187.9 | 186 |
| | | $A_g$ | 306.5 | 331 |
| | | $A_u$ | 436.2 | 423 |
| | | $B_u$ | 575.6 | 603 |

In order to investigate the stability of the structures, the Helmholtz free energy ($F$) of TiO phases vs. temperature ($T$) was drawn in Fig. 5(a). As it is seen, the vacancies make both c-TiO and m-TiO more stable, as the temperature increases. This is logically expected, since introducing oxygen vacancies enhances the entropy ($S$) in these compounds such that according to relation $F = U - TS$, it increases the stability of both c-TiO and m-TiO phases at higher temperatures. Moreover, the free energy diagram predicts that c-TiO is not more stable than m-TiO at high temperatures; this is in contrast with the experiment [6]. However, c-Ti$_{0.875}$O$_{0.875}$ is more stable than the m-TiO in the high temperatures that resolve this inconsistency. As is seen, m-TiO$_{1.2}$ and m-TiO$_{0.875}$ have the identical behavior with respect to temperature. This indicates that introducing either Ti or O vacancies makes m-TiO more stable as the temperature enhances. Additionally, the free energy diagram is able to represent structural phase transition from cubic into monoclinic phase bellow ~ 500 K. This value is lower than the experimental data (~ 1225 K) [11]. The value of the entropy for the c-Ti$_{0.875}$O$_{0.875}$ was presented in Table 4. The calculated entropy is closely according to the experiment [44]. The difference in structural phase transition and entropy value arises from ignoring the exact vacancy concentration discussed earlier and disordered vacancies in the cubic phase.

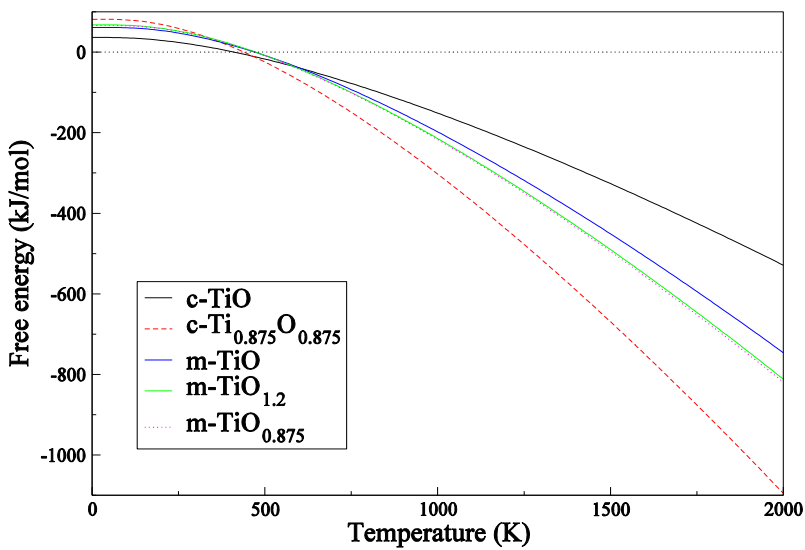

**Fig. 5:** The comparison of calculated free energy as a function of temperature for the c-TiO, c-Ti$_{0.875}$O$_{0.875}$, m-TiO, m-TiO$_{1.2}$ and m-TiO$_{0.875}$, respectively.

**Table 4.** Comparison between the calculated entropy (E.U. per mole) and experimental data of c-Ti$_{0.875}$O$_{0.875}$. The E.U. per mole is a Non-SI unit of molar entropy (1 E.U.= 4.184 J.K$^{-1}$.mol$^{-1}$).

|           | Temperature range (K) | Entropy (E.U./mole) |
|-----------|:---------------------:|:-------------------:|
| **This work** | 0-52 | 0.33 |
|           | 53-298 | 8.15 |
| **Exp. [44]** | 0-52 | 0.25 |
|           | 53-298 | 8.06 |

### 3.3 Prediction of pressure phase transition

In the earlier work, C. Leung *et al.* [7] predicted the pressure phase transition from monoclinic into cubic ~ 25 GPa using LDA and the damped version of the variable cell shape molecular-dynamics method for full relaxation of these structures. However, the experimental data illustrates this pressure as ~ 7.7 GPa [19]. This rather large discrepancy is attributed to the random ordering of the vacancies in the experimental samples of cubic structure. Therefore, it is attempted to reproduce the experimental data for the pressure phase transition from monoclinic into cubic structure by considering different configurations of vacant cubic structure, as was discussed in section. 3.1. Figure 6 demonstrates the plot of energy (per atom) vs. volume (E-V), and the corresponding formation enthalpy over the pressure for TiO phases. It is worthy to mention that the atomic positions were fully relaxed in each volume in E-V diagram.

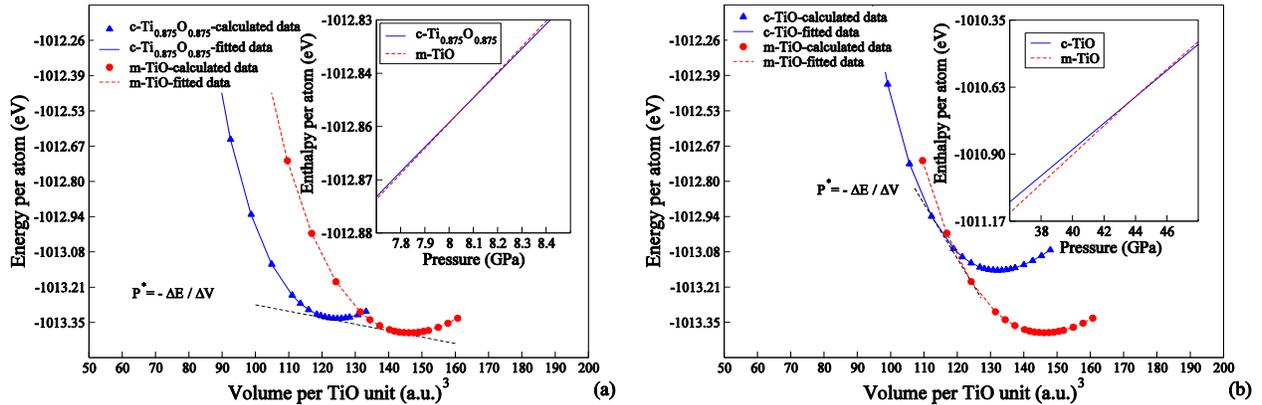

**Fig. 6.** The energy per atom vs. volume diagram using Birch–Murnaghan EOS for cubic and monoclinic phases (a) c-$Ti_{0.875}O_{0.875}$ and m-TiO, (b) vacancy-free c-TiO and m-TiO. $P^*$ is the pressure phase transition.

The results represents that the proposed structure for the cubic as c-$Ti_{0.875}O_{0.875}$ improves this discrepancy. The results depicted in onset of Fig. 6(a) shows that the pressure phase transition is ~ 8 GPa. The correction comes from the fact that introducing vacancies in the cubic phase increases the entropy of the system. Hence, the vacancies enhance the internal energy compared to the free-vacant cubic TiO. On the other hand, Ding $et$ $al$. [20] have recently observed that the structural transition occurs at 37.8 GPa. They have stated that the vacancy filling occurs by applying pressure. Thus, we have attempted to explain it by assuming that the cubic phase must be free from the vacancies. S. Bartkowski $et$ $al$. [8] have experimentally remarked that the vacancy-free cubic samples can be prepared by annealing a finely ground powder of monoclinic TiO at a high temperature and pressure. Almost all the vacancies can be eliminated from the monoclinic phase at above 8 GPa. Using the vacancy-free rock-salt TiO structure (c-TiO, stable above ~ 10 GPa) changes the pressure phase transition to ~ 44 GPa as displayed in onset of Fig. 6 (b). Additionally, Andersson $et$ $al$. [30] thermodynamically showed that the cubic phase becomes more stable than the monoclinic phase above 40 GPa. This result is rather recovered in the present calculation. Based on the results, the role of vacancy sites and their concentration are key parameters in determining the pressure phase transition of $TiO_x$ compounds.

## 4 Conclusions

In the present study, the structural, electronic and dynamical properties of $TiO_x$ in two phases of cubic ($Fm\overline{3}m$) and monoclinic ($A2/m$) structures were investigated using density functional theory calculations. The structural calculations reveal that the vacancy defect lowers the bulk modulus in TiO phases. The electronic structure of the $TiO_x$ compounds shows the metallic behavior of these structures in such a way that the conduction states below the Fermi level mainly originate from Ti-Ti bonds. The observed sharp peaks in the conduction bands coming from Ti-$3d$ orbitals of defected $TiO_x$ structures are driven by the vacancy sites. Moreover, an energy gap is appeared between Ti-$3d$ and O-$2p$ orbitals of $TiO_x$ below the Fermi level, which may vary by the vacancy concentration. The phonon calculations represent the dynamical instability of c-TiO at room temperature. Including 12.5 % vacancies for both Ti and O in this phase (c-$Ti_{0.875}O_{0.875}$) greatly correct the imaginary modes. However, the c-TiO can be stabilized at above ~ 10 GPa, which is consistent with the experimental data. In addition, considering O or Ti vacancies in m-TiO reveled that m-$TiO_{0.833}$ is more stable than m-$TiO_{1.2}$. Investigating the phonon dispersion curves of $TiO_x$ compounds indicate a phonon gap in the optical bands which originates from the mass difference of Ti and O. Moreover, the absence of this gap between the acoustic and optical bands suggests that $TiO_x$ compounds are not good candidates for thermal conductivity. The calculated free energy vs. temperature cleared up c-$Ti_{0.875}O_{0.875}$ and m-TiO phases are stable in a higher and lower temperature, respectively. The results obtained by fitting E-V data to Birch–Murnaghan EOS indicated that the monoclinic phase is more stable than the cubic one at lower pressures. Furthermore, the structural phase transition from monoclinic to cubic TiO under pressure was studied by modifying the vacancy concentration. The transition from the monoclinic into the defected cubic TiO (c-$Ti_{0.875}O_{0.875}$) occurred at the pressure ~ 8 GPa, whereas this pressure was ~ 44 GPa during the transition from monoclinic into the vacancy-free cubic TiO (c-TiO). The obtained results well describe the experimental data. Totally, it can be concluded that engineering both vacancy sites and their concentration are the key parameters in fabricating the desired Ti-O base systems.

## Acknowledgements

Partial financial support by the Research Council of the University of Tehran is acknowledged. The computation in this work was partially performed using the facilities of the Center for Information Science in JAIST.

# References

[1]     S. Srivastava, J. P. Thomas, and K. T. Leung, Nanoscale **11**, 18159 (2019).

[2]     M.-Y. Tse, X. Wei, and J. Hao, Physical Chemistry Chemical Physics **18**, 24270 (2016).

[3]     L. Mohrhusen, J. Kräuter, and K. Al-Shamery, Physical Chemistry Chemical Physics (2021).

[4]     P. Pai, F. K. Chowdhury, T.-V. Dang-Tran, and M. Tabib-Azar, in *SENSORS, 2013 IEEE* (IEEE, 2013), pp. 1.

[5]     K. Li, J. Wang, and A. R. Oganov, The Journal of Physical Chemistry Letters **12**, 5486 (2021).

[6]     D. Watanabe, J. Castles, A. Jostsons, and A. Malin, Acta Crystallographica **23**, 307 (1967).

[7]     C. Leung, M. Weinert, P. B. Allen, and R. M. Wentzcovitch, Physical Review B **54**, 7857 (1996).

[8]     S. Bartkowski, M. Neumann, E. Kurmaev, V. Fedorenko, S. Shamin, V. Cherkashenko, S. Nemnonov, A. Winiarski, and D. Rubie, Physical Review B **56**, 10656 (1997).

[9]     I. Martev, Vacuum **58**, 327 (2000).

[10]    K. Grigorov, G. Grigorov, L. Drajeva, D. Bouchier, R. Sporken, and R. Caudano, Vacuum **51**, 153 (1998).

[11]    S. P. Denker, Journal of Applied Physics **37**, 142 (1966).

[12]    N. Doyle, J. Hulm, C. Jones, R. Miller, and A. Taylor, Physics Letters A **26**, 604 (1968).

[13]    D. Wang, C. Huang, J. He, X. Che, H. Zhang, and F. Huang, ACS omega **2**, 1036 (2017).

[14]    T.-H. Kim, M.-H. Kim, S. Bang, D. K. Lee, S. Kim, S. Cho, and B.-G. Park, IEEE Transactions on Nanotechnology **19**, 475 (2020).

[15]    W. Banerjee, Q. Liu, H. Lv, S. Long, and M. Liu, Nanoscale **9**, 14442 (2017).

[16]    H. Zhang, C. Cheng, H. Zhang, R. Chen, B. Huang, H. Chen, and W. Pei, Physical Chemistry Chemical Physics **21**, 23758 (2019).

[17]    A. Palagushkin, D. Roshchupkin, F. Yudkin, D. Irzhak, O. Keplinger, and V. Privezentsev, Journal of Applied Physics **124**, 205109 (2018).

[18]    S. Prada, M. Rosa, L. Giordano, C. Di Valentin, and G. Pacchioni, Physical Review B **83**, 245314 (2011).

[19]    D. Watanabe, O. Terasaki, A. Jostsons, and J. Castles, North-Holland Publishing Co, 238 (1970).

[20]    J. Ding, T. Ye, H. Zhang, X. Yang, H. Zeng, C. Zhang, and X. Wang, Applied Physics Letters **115**, 101902 (2019).

[21]    P. Giannozzi *et al.*, Journal of physics: Condensed matter **21**, 395502 (2009).

[22]    D. Vanderbilt, Physical review B **41**, 7892 (1990).

[23]    J. P. Perdew, K. Burke, and M. Ernzerhof, Physical review letters **77**, 3865 (1996).

[24]    H. J. Monkhorst and J. D. Pack, Physical review B **13**, 5188 (1976).

[25]    J. D. Head and M. C. Zerner, Chemical physics letters **122**, 264 (1985).

[26]    A. Togo and I. Tanaka, Scripta Materialia **108**, 1 (2015).

[27]    K. Parlinski, Z. Li, and Y. Kawazoe, Physical Review Letters **78**, 4063 (1997).

[28]    J. Graciani, A. Márquez, and J. F. Sanz, Physical Review B **72**, 054117 (2005).

[29]    M. Banus, T. Reed, and A. Strauss, Physical Review B **5**, 2775 (1972).

[30]    D. A. Andersson, P. A. Korzhavyi, and B. Johansson, Physical Review B **71**, 144101 (2005).

[31]    K. Momma and F. Izumi, Journal of applied crystallography **44**, 1272 (2011).

[32]    V. Ern and A. Switendick, Physical Review **137**, A1927 (1965).

[33]    A. Neckel, P. Rastl, R. Eibler, P. Weinberger, and K. Schwarz, Journal of Physics C: Solid State Physics **9**, 579 (1975).

[34]    G. Hobiger, P. Herzig, R. Eibler, F. Schlapansky, and A. Neckel, Journal of Physics: Condensed Matter **2**, 4595 (1990).

[35]    J. Schoen and S. Denker, Physical Review **184**, 864 (1969).

[36]    S. Barman and D. Sarma, Physical Review B **49**, 16141 (1994).

[37]    M. Kostenko, A. Lukoyanov, V. Zhukov, and A. Rempel, Journal of Solid State Chemistry **204**, 146 (2013).

[38]    S. Baroni, S. De Gironcoli, A. Dal Corso, and P. Giannozzi, Reviews of modern Physics **73**, 515 (2001).

[39]    K. Schwarz and P. Blaha, Computational Materials Science **28**, 259 (2003).

[40]    G. Raunio, L. Almqvist, and R. Stedman, Physical Review **178**, 1496 (1969).

[41]    U. Argaman, R. E. Abutbul, Y. Golan, and G. Makov, Physical Review B **100**, 054104 (2019).

[42]    M. T. Dove, American Mineralogist **82**, 213 (1997).

[43]    G. Hasegawa, T. Sato, K. Kanamori, K. Nakano, T. Yajima, Y. Kobayashi, H. Kageyama, T. Abe, and K. Nakanishi, Chemistry of Materials **25**, 3504 (2013).

[44]    C. H. Shomate, Journal of the American Chemical Society **68**, 310 (1946).